# The Effect of Empathic Expression Levels in Virtual Human Interaction: A Controlled Experiment

Sung Park¹*, Daeho Yoon¹, Jungmin Lee²
*Corresponding author: Sung Park (email: sjp@taejae.ac.kr)

¹ Human-Centric AI Center, Taejae University
² Department of Human-Centered AI, Sangmyung University

**Abstract**
As artificial intelligence (AI) systems become increasingly embedded in everyday life, the ability of interactive agents to express empathy has become critical for effective human-AI interaction, particularly in emotionally sensitive contexts. Rather than treating empathy as a binary capability, this study examines how different levels of empathic expression in virtual human interaction influence user experience. We conducted a between-subject experiment (n = 70) in a counseling-style interaction context, comparing three virtual human conditions: a neutral dialogue-based agent, a dialogue-based empathic agent, and a video-based empathic agent that incorporates users' facial cues. Participants engaged in a 15-minute interaction and subsequently evaluated their experience using subjective measures of empathy and interaction quality. Results from analysis of variance (ANOVA) revealed significant differences across conditions in affective empathy, perceived naturalness of facial movement, and appropriateness of facial expression. The video-based empathic expression condition elicited significantly higher affective empathy than the neutral baseline ($p < .001$) and marginally higher levels than the dialogue-based condition ($p < .10$). In contrast, cognitive empathy did not differ significantly across conditions. These findings indicate that empathic expression in virtual humans should be conceptualized as a graded design variable, rather than a binary capability, with visually grounded cues playing a decisive role in shaping affective user experience.

**Keywords**
Empathic virtual agents, virtual humans, multimodal interaction, affective empathy, facial expression, human-AI interaction.

## 1. Introduction

Empathy acts as a fundamental glue of human social interaction, facilitating trust, understanding, and emotional regulation (Decety & Lamm, 2006). As virtual agents and robots increasingly assume roles in healthcare, education, and companionship (Lukasik, 2025; Masala & Giorgi, 2025), the ability to perceive and respond to human emotions, often referred to as artificial empathy (Asada, 2015), has become an important research focus. While early conversational agents relied primarily on textual inputs, advances in affective computing have underscored the importance of multimodal approaches that more closely reflect the complexity of human communication (D'Mello & Kory, 2015). Rather than assuming empathy as a binary capability, this study examines how different levels of empathic expression in virtual human interaction influence user experience in emotionally sensitive contexts.

Research on artificial empathy commonly distinguishes between cognitive empathy (i.e., understanding another's perspective) and affective empathy (i.e., responding to or sharing another's emotional state). Park and Whang (2022) argue that for artificial entities to be perceived as truly empathic, they must integrate these multi-dimensional aspects into their behavioral design. Paiva et al. (2017) emphasize that effective virtual agents should exhibit both dimensions through congruent verbal and non-verbal behaviors. In counseling contexts, where perceived empathy plays a central role in forming a therapeutic alliance, virtual agents have shown potential for supportive interaction (Lisetti et al., 2013). However, prior findings also suggest that the effectiveness of such agents strongly depends on how accurately and appropriately they respond to users' emotional states.

Affective computing research has explored emotion recognition using a variety of physiological and behavioral signals, including brain activity and heart rate dynamics (Marín-Morales et al., 2018). Among these modalities, facial expressions remain one of the most immediate and natural channels for emotional communication. Advances in computer vision and deep learning have substantially improved facial expression recognition (FER), with Convolutional Neural Networks (CNNs) achieving strong performance in decoding both static and dynamic facial cues (Li & Deng, 2020). More recent architectures, such as EmoNet, enable the estimation of continuous valence and arousal in naturalistic conditions, capturing subtle affective changes beyond categorical emotion labels (Toisoul et al., 2021). Prior work further suggests that integrating multiple modalities can enhance affective understanding, as multimodal approaches often outperform unimodal ones in emotionally sensitive applications (Sangeetha et al., 2024).

Despite these technological advances, challenges remain in understanding how empathic signals should be expressed by virtual humans to foster positive user experience. Existing systems often rely either on dialogue-based sentiment analysis or on predefined non-verbal behaviors, making it difficult to isolate the specific contribution of different empathic expression strategies. As noted by Potdevin et al. (2021), misalignment between verbal content and non-verbal expression can undermine perceived intimacy and trust. Techniques such as FACS-based expression modeling (Cohn et al., 2007; Ekman & Friesen, 1983) and continuous valence–arousal tracking (Garg & Verma, 2020) provide mechanisms for generating expressive behavior, yet empirical evaluations comparing different levels of empathic expression remain limited.

To address this gap, the present study treats empathic expression level as a key design variable in virtual human interaction. We examine three virtual human conditions that differ in empathic expression: a neutral dialogue-based baseline, a dialogue-based empathic agent, and a video-based empathic agent that incorporates users' facial cues during interaction. Through a controlled between-subject experiment conducted in a counseling-style interaction scenario, we evaluate how these differing levels of empathic expression influence users' perceived empathy and interaction quality. By systematically comparing verbal-only and visually augmented empathic expressions, this work aims to clarify when increased expressiveness meaningfully enhances user experience, rather than assuming that greater technical sophistication alone leads to better outcomes. Beyond demonstrating the benefits of multimodal empathy, this study contributes by empirically disentangling levels of empathic expression in virtual humans and examining their differential effects on affective and cognitive dimensions of perceived empathy. By treating empathic expression as a graded design variable, rather than a monolithic system capability, this work offers design-relevant insights for the development of empathic virtual agents. This study addresses the following research questions:

RQ1: Does increasing the level of empathic expression in virtual humans enhance users' affective empathy?
RQ2: Does video-based empathic expression lead to higher perceived interaction quality compared to dialogue-based empathy?

## 2. Method

*2.1 Participants*

A total of 70 participants (27 males, 43 females) were recruited for this study. Participants were university students and staff recruited via online community boards. An a priori power analysis using G*Power indicated that a sample size of 66 was required to detect a medium effect size (0.4) with a power of .80 and an alpha level of .05 using an F-test for a one-way ANOVA with three groups. To account for potential data loss, we recruited 70 participants. They were randomly assigned to one of three experimental conditions: the neutral dialogue-based baseline (n = 23), the dialogue-based empathic agent condition (n = 24), and the video-based empathic agent condition (n = 23).

*2.2 Design and Procedure*

The study employed a between-subjects experimental design. The independent variable was the level of empathy expression utilized by the virtual human (three levels). The dependent variables included affective empathy, cognitive empathy, perceived naturalness of facial movement, and appropriateness of facial expression. The study protocol was reviewed and approved by the Institutional Review Board (IRB) of Sangmyung University (IRB-SMU-S-2023-3-012), and was classified as minimal risk.

Upon arrival at the laboratory, participants were briefed on the experimental procedure and provided informed consent (see Figure 1). The experiment began with a familiarization phase lasting approximately 10 minutes, during which participants interacted with the virtual human interface to become comfortable with the speech recognition and response dynamics. Following this, participants engaged in a 15-minute interaction with the virtual human. The interaction context was a counseling-style scenario focused on romantic relationship concerns, selected to elicit a wide range of emotional responses and engage participants in a personally relevant dialogue. After the interaction concluded, participants completed a post-experiment questionnaire measuring various psychological and technical constructs on a 7-point Likert scale (1 = strongly disagree, 7 = strongly agree).

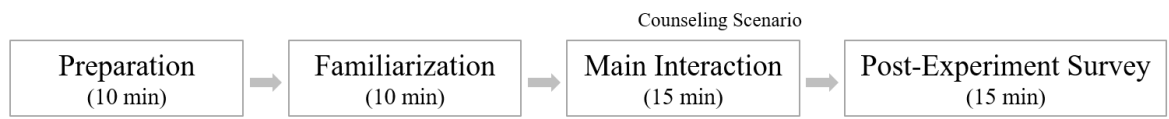

Figure 1. Overview of the experimental procedure.

*2.3 Empathic Expression Conditions*

In this study, empathic expression level refers to the extent to which a virtual human externally manifests empathy through verbal and non-verbal cues, ranging from neutral expression to multimodally grounded affective mirroring. The virtual human system was developed with three distinct levels of capability (see Table 1). All conditions utilized Naver Clova's Automatic Speech Recognition (ASR) and Text-to-Speech (TTS) technologies for verbal interaction. Across all conditions, the linguistic content, dialogue structure, and response appropriateness were held constant to isolate the effect of non-verbal empathic expression.

Table 1.
Design of the Experimental Conditions with Different Levels of Empathic Expression

| Condition | Description | Participant Emotion Recognition | | Virtual Human Empathic Expression | |
|---|---|---|---|---|---|
| | | Facial | Speech | Facial Expression | Speech |
| Condition 1 (baseline) | Neutral dialogue based expression | X | O | X (neutral expression) | LLM-based |
| Condition 2 | Dialogue-based empathy | X | O | O (derived from dialogue sentiment) | LLM-based |
| Condition 3 | Video-based empathy | O | O | O (derived from participant's facial cues and dialogue) | LLM-based |

Note. O indicates the presence of a modality or expressive feature, whereas X indicates its absence.

Condition 1 - Neutral dialogue-based expression (baseline): in this condition, the virtual human maintained a neutral facial expression throughout the interaction. The agent utilized a dialogue generation model to produce supportive dialogue responses.

Condition 2 - Dialogue-based empathy: in this condition, the virtual human's facial expressions were driven solely by the sentiment of its own verbal responses. The system performed sentiment analysis on the generated text using a transformer-based large language model (GPT-3.5 Turbo) to determine the appropriate emotional tone (valence and arousal). These values were mapped to facial expressions, ensuring the agent's face matched the content of its speech (e.g., looking sad when offering condolences). The emotional tone of each utterance was represented along continuous valence and arousal dimensions, following predefined scale criteria (see Table 2).

| Dimension | Value Range | Interpretation |
|---|---|---|
| Valence | -1.0 | Extremely negative emotion (extreme sadness) |
| | -0.8 | Strongly negative emotion (sadness) |
| | -0.5 | Mildly negative emotion (discomfort) |
| | 0.0 | Neutral emotion |
| | 0.5 | Positive emotion (satisfaction) |
| | 0.8 | Strongly positive emotion (happiness) |
| | 1.0 | Extremely positive emotion (ecstatic happiness) |
| Arousal | -1.0 | Extremely low arousal (very bored) |
| | -0.5 | Low arousal (slightly bored) |
| | 0.0 | Neutral arousal |
| | 0.5 | High arousal (excited) |
| | 1.0 | Extremely high arousal (highly excited) |

Table 2.
Valence and arousal scale definitions used for empathic expression generation

Condition 3 - Video-based empathy: this condition represented the most expressive form of empathic expression in the study. It integrated real-time facial expression recognition of the user with the dialogue system. The user's facial video feed was analyzed using the EmoNet architecture (Toisoul et al., 2021) to estimate continuous valence and arousal levels. The system then synthesized this input with the sentiment of the conversation to generate congruent facial expressions using the Facial Action Coding System (FACS) (Ekman & Friesen, 1983). To determine the final expressive state of the virtual human, basic facial action units were integrated within a continuous valence–arousal space to derive dominant emotional expressions (see Figure 2).

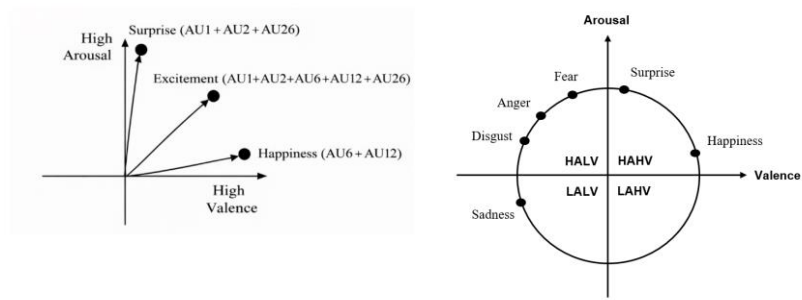

Figure 2. Mapping and Integration of Facial Action Units into Valence–Arousal Emotion Space

The 3D virtual character's face was animated via BlendShape manipulation, dynamically adjusting facial expressions based on continuous valence-arousal estimates (High Arousal High Valence, High Arousal Low Valence, Low Arousal High Valence, Low Arousal Low Valence).

## 3. Measures
After the interaction, participants completed a post-experiment questionnaire consisting of 16 items assessing their subjective experience with the virtual human. All items were rated on a 7-point Likert scale (1 = strongly disagree, 7 = strongly agree). Affective empathy was assessed using items capturing participants' emotional resonance with the virtual human, including feelings of intimacy, liking, and emotional connection (e.g., "I felt emotionally close to the virtual human"). Cognitive empathy measured the extent to which participants perceived the virtual human as understanding their thoughts and feelings (e.g., "I felt that the virtual human understood my thoughts and emotions"). Interaction quality was evaluated through items assessing the perceived naturalness and appropriateness of the virtual human's behavior, including facial movement naturalness, facial expression appropriateness, dialogue appropriateness, and speech naturalness. In addition, several items measured relational outcomes and user attitudes, such as trust in the virtual human, enjoyment of the interaction, satisfaction with the counseling experience, and intention to engage in future interactions with the virtual human. The full list of questionnaire items is provided in the Appendix.

## 4. Results and Analysis
*4.1 Overview of Statistical Analysis*
To examine the effect of empathic expression level on user experience, a one-way analysis of variance (ANOVA) was conducted on post-interaction questionnaire responses. The independent variable was the level of empathic expression of the virtual human with three conditions: neutral dialogue-based expression (baseline), dialogue-based empathy, and video-based empathy. The dependent variables consisted of 16 subjective evaluation items measuring perceived empathy and interaction quality. Post-hoc comparisons were conducted using adjusted p-values.

*4.2 Affective and Cognitive Empathy*
Significant differences were observed across conditions for affective empathy. As shown in Figure 3(a), participants reported significantly higher affective empathy in the video-based empathy condition compared to the neutral baseline condition (p < .001). The dialogue-based empathy condition also showed higher affective empathy than the baseline condition, reaching marginal significance (p < .10). In addition, affective empathy ratings in the video-based condition were marginally higher than those in the dialogue-based condition (p < .10), indicating a graded effect of empathic expression level.

In contrast, no statistically significant differences were found across conditions for cognitive empathy (Figure 3(b)). Although the video-based empathy condition demonstrated higher mean scores than the baseline condition, this difference did not reach statistical significance (p = .111). These results suggest that while enhanced empathic expression strongly influences affective resonance, it has a more limited effect on users' cognitive evaluations of being understood. Cognitive empathy may rely more heavily on linguistic coherence and semantic understanding, which were held constant across conditions, whereas affective empathy appears to be more sensitive to non-verbal and visual empathic cues.

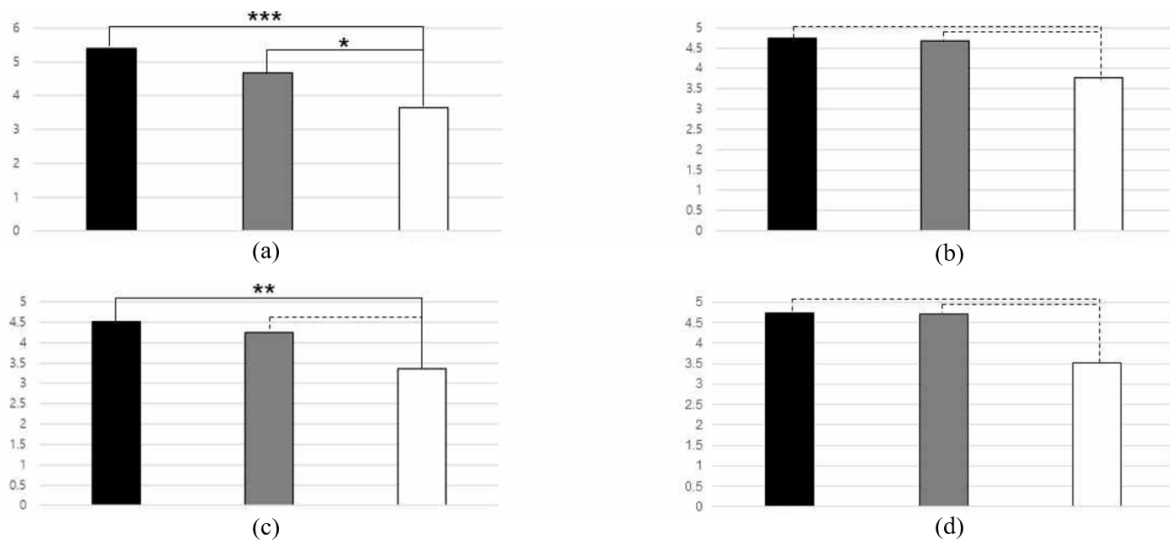

Figure 3. Mean ratings for selected subjective measures that showed significant or near-significant differences across empathic expression conditions. Black bars represent the video-based empathic expression condition, gray bars represent the dialogue-based empathic expression condition, and white bars represent the neutral dialogue-based baseline condition. (a) perceived emotional similarity, (b) perceived cognitive understanding, (c) naturalness of facial movement, and (d) appropriateness of facial expression. Error bars indicate standard errors. Statistical significance is denoted as *p* < .10 (*), *p* < .05 (**), p < .001 (***).

*4.3 Facial Movement Naturalness and Expression Appropriateness*
Perceived naturalness of facial movement differed significantly across empathic expression levels (Figure 3(c)). Post-hoc analysis revealed that only the video-based empathy condition elicited significantly higher ratings compared to the neutral baseline condition (p < .05). No significant difference was observed between the dialogue-based empathy and baseline conditions.

A similar pattern emerged for perceived facial expression appropriateness (Figure 3(d)). Both the video-based empathy and dialogue-based empathy conditions showed higher ratings than the neutral baseline, with marginal significance (p < .10). Among the three conditions, the video-based empathy condition consistently exhibited the highest mean ratings, indicating that visually grounded empathic expression contributed most strongly to perceived expressiveness.

*4.4 Summary of Key Empathic Interaction Indicators*

Taken together, the results indicate that the level of empathic expression plays a critical role in shaping user experience with virtual humans. Video-based empathic expression, which integrates users' facial cues into the agent's non-verbal behavior, significantly enhanced affective empathy and non-verbal interaction quality. While cognitive empathy did not differ significantly across conditions, the observed trends suggest that multimodal empathic expression may contribute to a more comprehensive perception of empathy when combined with dialogue-based strategies. These findings support the notion that affective and non-verbal dimensions of empathy are particularly sensitive to visual-level empathic cues, and highlight video-based empathic expression as a key indicator of empathic interaction quality.

## 5. Discussion and Conclusion

The present study aimed to evaluate the effectiveness of a video-based empathic virtual human that integrates multimodal cues for romantic relationship counseling. The results provide strong evidence that incorporating real-time facial expression recognition into the behavioral model of a virtual human significantly enhances its perceived affective empathy and naturalness.

The superiority of the video-based condition in eliciting affective empathy supports the theoretical framework that empathy is a multimodal construct requiring the synchronization of internal understanding and external behavioral mimicry (Milcent et al., 2022). By utilizing EmoNet (Toisoul et al., 2021) to track the user's emotional state continuously, the system was able to generate "congruent" facial expressions, mirroring the user's valence and arousal in real-time. This mechanism likely fostered a sense of "being seen" and emotional validation, which is crucial in counseling contexts (Lisetti et al., 2013). The fact that the dialogue-based condition fell short of the video-based condition suggests that verbal empathy alone, even when augmented with sentiment-driven expressions, is less effective than a system that actively attends to the user's non-verbal signals.

The findings also validate the utility of dimensional emotion models (Garg & Verma, 2020) in driving virtual human animation. The successful differentiation in facial movement naturalness suggests that the blend-shape manipulation based on the four quadrants of valence and arousal resulted in smoother and more contextually appropriate animations than static or text-triggered expressions.

However, several limitations were identified. Qualitative feedback indicated that while the facial expressions were effective, users desired more subtle non-verbal behaviors such as breathing movements and small gestures to enhance realism. Additionally, the current ASR verification process disrupted the immersion, pointing to a need for more robust, seamless speech processing. Future research should focus on integrating these physiological cues (e.g., Heart Rate Variability) as suggested by Marín-Morales et al. (2018), and refining the conversation scenarios to encourage deeper emotional disclosure. This study presents a significant step forward in the development of empathic AI. By successfully integrating video-based facial expression recognition with generative dialogue, we created a virtual human capable of multimodal empathy that users perceive as more natural and emotionally resonant. These findings have broad implications for the design of digital health interventions, suggesting that the "eyes" of the AI—its ability to see and mirror the user—are just as important as its "voice." As we continue to refine these systems, the potential for AI to provide meaningful, accessible mental health support becomes increasingly tangible. Taken together, this study demonstrates that empathic expression in virtual humans should be treated as a graded design variable, in which visually grounded non-verbal cues play a critical role in shaping affective user experience, beyond verbal empathy alone.

**Acknowledgements**
This research was supported by Basic Science Research Program through the National Research Foundation of Korea(NRF) funded by the Ministry of Education(RS-2022-NR075075). The experimental system utilized language model resources provided by Scatter Lab.

**Appendix A. Questionnaire Items**
Note. All items were originally administered in Korean and translated into English for reporting purposes. Participants rated each item on a 7-point Likert scale (1 = strongly disagree, 7 = strongly agree).

*Affective and Cognitive Empathy*
1. I felt a sense of intimacy with the virtual human.
2. I felt empathy toward the virtual human.
3. I found the virtual human appealing.
4. I trusted what the virtual human said.
5. I enjoyed the conversation with the virtual human.
6. I was satisfied with the romantic counseling interaction with the virtual human.
7. I would like to engage in romantic counseling with the virtual human again in the future.
8. I felt that the virtual human experienced emotions similar to mine.
9. I felt that the virtual human understood my thoughts and emotions.

*Interaction Quality*
10. The appearance of the virtual human was realistic.
11. The virtual human's facial movements were natural.
12. The virtual human's facial expressions were appropriate.
13. The virtual human's voice sounded natural.
14. The virtual human's responses were appropriate.

*Immersion*
15. I felt that the virtual human and I were present in the same space.
16. I felt immersed in the interaction with the virtual human.